# Working Papers in Computer Science

# Towards a Semantic Model of the GDPR Register of Processing Activities

## Dublin City University

*Paul Ryan, Dublin City University*
*Dr. Harshvardhan J. Pandit, Trinity College Dublin*
*Dr. Rob Brennan, Dublin City University*

# Towards a Semantic Model of the GDPR Register of Processing Activities

Paul Ryan
ADAPT, School of Computing
Dublin City University
Glasnevin, Dublin 9, Ireland
paul.ryan76@mail.dcu.ie

Harshvardhan J. Pandit
ADAPT Centre
Trinity College, Dublin, Ireland
pandith@tcd.ie

Rob Brennan
ADAPT, School of Computing
Dublin City University
Glasnevin, Dublin 9, Ireland
rob.brennan@dcu.ie

## ABSTRACT

A core requirement for GDPR compliance is the maintenance of a register of processing activities (ROPA). Our analysis of six ROPA templates from EU data protection regulators shows the scope and granularity of a ROPA is subject to widely varying guidance in different jurisdictions. We present a consolidated data model based on common concepts and relationships across analysed templates. We then analyse the extent of using the Data Privacy Vocabulary - a vocabulary specification for GDPR. We show that the DPV currently does not provide sufficient concepts to represent the ROPA data model and propose an extension to fill this gap. This will enable creation of a pan-EU information management framework for interoperability between organisations and regulators for GDPR compliance.

## Introduction

A Register of Processing Activities (ROPA) is a comprehensive record of the personal data processing activities of an organisation and is a central document in the GDPR compliance processes. ROPA's are important in the GDPR compliance process and enable organisations to demonstrate the principle of accountability [1]. The contents of a ROPA as required by Article 30 of GDPR must contain information necessary to describe the identity of the data controller and a description of the personal data processing in terms of purposes, data subjects, personal data, recipients, data transfers, and technical and organisational measures in place. A ROPA record may be documented through paper or digital mediums and must be made available to a regulatory authority for inspection upon request to audit. The creation of a ROPA can be quite a daunting and arduous task involving the discovery, recording and documenting of all business processes pertaining to processing of personal data. The International Association of Privacy Professionals (IAPP) reported the primary approach to maintaining ROPA's by organisations were informal tools, such as email, manual spreadsheets and in person communication [2]. Several EU data protection regulators have provided templates to assist organisations in maintaining ROPAs. As we demonstrate in this paper - even though ROPAs share a common requirement in the form of Article 30 of GDPR, templates from regulators vary in their complexity, scope, and granularity of information required.

We propose to consolidate these different ROPA requirements into a common semantic data model to assist regulators and organisations with the GDPR compliance process across jurisdictions. For this, we evaluate the extent of the Data Privacy Vocabulary (DPV) [3] - a community vocabulary specification for GDPR - to represent the information required by ROPA templates from data protection authorities. Our analysis shows the DPV is currently missing important concepts in relation to ROPAs and GDPR compliance. We identify the extent of work required to fill this gap towards a semantic model of ROPAs.

This paper thus provides an insight into common elements and variations in recommended ROPAs across the EU and shows how the DPV may be applied for creating a semantic model for use in information management technologies. A common ROPA model greatly assists organisations and authorities in the GDPR compliance process - especially when it spans multiple jurisdictions. Our approach is the first step to facilitate use of ROPAs in regulatory compliance and enable developments such as process automation, the digitising of data, the use of semantic methods and machine learning algorithms [4]. In addition to these, the semantic model of ROPA enables use of queries to identify areas of non-compliance, and remedy or mitigate accordingly, as part of the accountability framework for an organisation [5].

## Analysis of ROPA's

We analysed websites of EU data protection regulatory authorities and identified 14 ROPA templates. Our current analysis focuses on 6 of these based on their use of English language. We have identified that these ROPA templates differ greatly in extent, where some ROPA templates have as little as 12 data input fields whilst others have up to 34 fields. Our findings indicate that all 6 ROPA templates meet the minimum

Article 30 requirements, and that some of the ROPA templates were a direct transcribed version of the GDPR legislation whereas others contained additional information requirements. This is due to the differing perspectives of data protection regulators. The French Data Protection regulator (CNIL) has provided a substantial ROPA template to assist organisation in transforming the ROPA into a GDPR compliance tool [6]. The CNIL recommends gathering all details related to the personal data processing of an organisation in one document to simplify compliance of data protection rules and to help identify actions that the organisation needs to take [6]. Similarly, ICO - the data protection regulator for the UK – provides additional data fields beyond the mandatory required data fields towards assisting Data Protection Officers in fulfilling their duties. Therefore, in practice the use of ROPA goes beyond its origins in Article 30 towards meeting larger compliance requirements such as the accountability principle of the GDPR. In the first step in our analysis, we identified and consolidated data fields across 6 English language ROPA templates based on their relation to GDPR clauses and common or unique information input requirements. Through this process we identified 43 unique data input fields.

## A Semantic Model of ROPA

In the second stage of our analysis, we reviewed the 43 unique data entry fields from the ROPA templates to establish their relationship to GDPR concepts. These concepts were mapped to a semantic model as displayed in Figure 1.

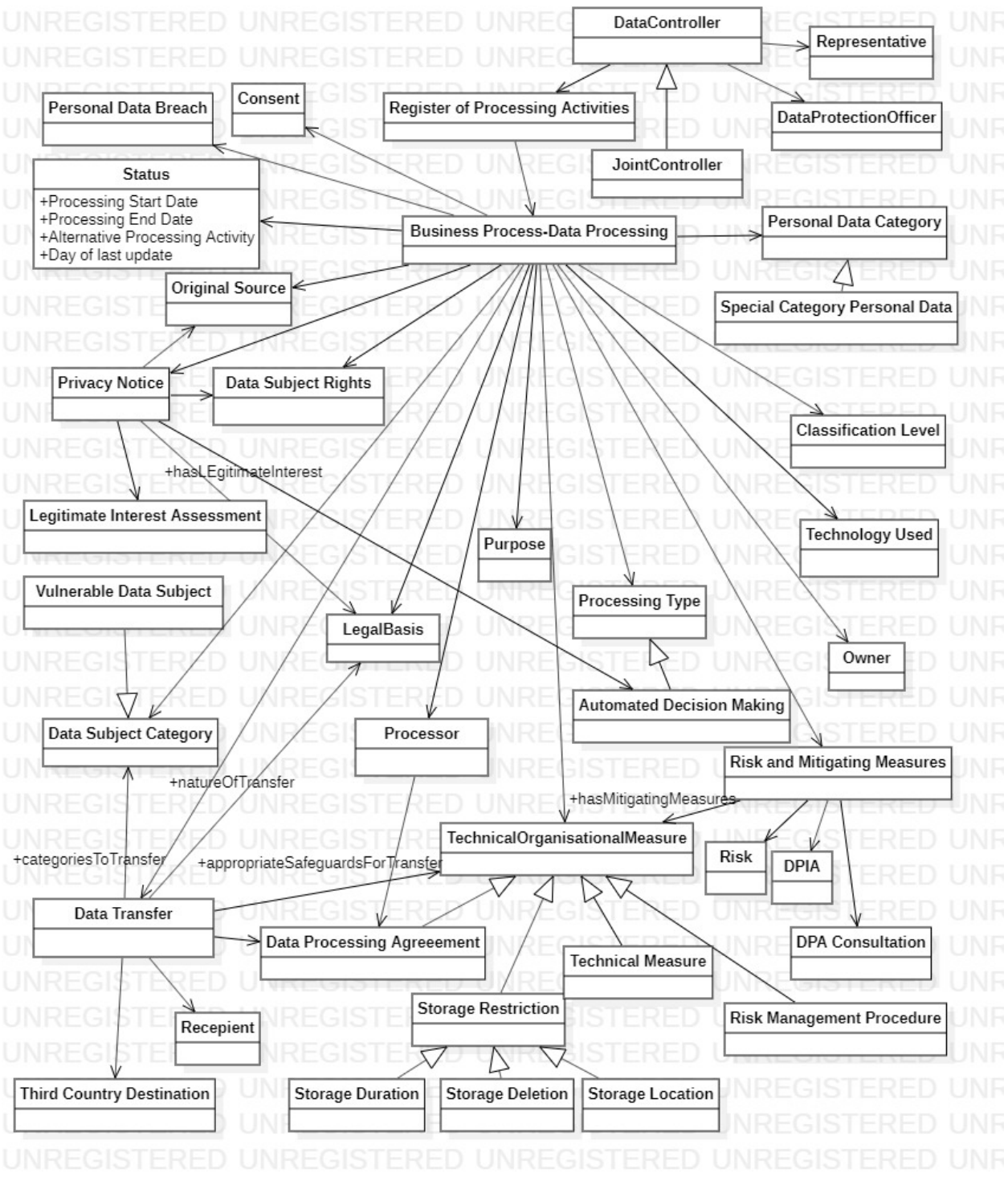

Fig 1 Combined ROPA Model Derived from DPA Templates and GDPR

## Mapping a ROPA using the DPV

We identified the Data Privacy Vocabulary (DPV) [3] as the most relevant and suitable resource to map our ROPA due to its status as a community specification through the W3C Data Privacy Vocabulary and Controls Community Group (DPVCG). The DPVCG was set up to develop a community and standardize vocabularies towards interoperability in the context of data privacy laws such as the GDPR. Our next step consisted of an evaluation of the extent that the DPV could be used to represent the 43 unique GDPR concepts gathered in our consolidated ROPA template. For this, we mapped relevant GDPR concepts to the DPV and categorised the mapping based on the semantics of information as: "Exact" if the field exactly corresponds to an existing DPV concept indicating no change required, 'Partial' if the data field as a corresponding concept in the DPV that needs to be extended or supplemented with properties, 'Complex/Partial' if the required field can be specified using a combination of multiple concepts in DPV, and 'None' if the concept is missing and needs to be added to the DPV [7].

In Table 1 we provide a detailed results table for the mapping of each GDPR concept showing the relative DPV match, the DPV mapping outcome, and the templates that the field is present in. The study shows that the DPV goes some way towards a full mapping, but it requires additional concepts to be added. A summary of the mapping of the GDPR concept with DPV is displayed in table 2.

Table 2 Summary of status of Match GDPR concept to DPV

| Match Status | Number of GDPR Concepts |
|---|---|
| Exact | 14 |
| Partial | 15 |
| Complex/Partial | 3 |
| None | 11 |
| Total | 43 |

In its current format, our study identifies that the DPV requires a number of additional GDPR concepts to be added to fully map these ROPA templates. Among the 11 additional concepts required are International Transfers, Controller Name and Contact Details, Original Source of Data, Data Protection Officer, Data Protection Impact Assessment, Data Subject Rights, Risk, Privacy Notice, Representative & Data Breach (Refer to Table 1 for a full list, under column DPV matching status)

Table 1 Mapping Table ROPA GDPR concept to DPV

| GDPR Regulation | Regulator Template GDPR Concept | Art.30 Mandatory | Related DPV Concept | DPV mapping outcome | Combined No. of Specified Field Values vs DPV | Belgium (34) | Cyprus (12) | Denmark (12) | Finland (13) | Luxembourg (14) | UK (33) |
|---|---|---|---|---|---|---|---|---|---|---|---|
| 30 | Register of Processing Activities | Y | No DPV Concept | None | | Y | Y | Y | Y | Y | Y |
| 30(1)(a) | DataController | Y | dpv:DataController | Exact | | Y | Y | Y | Y | Y | Y |
| 30(1)(a) | Controller name and contact details | Y | Many suitable vocabularies | None | | Y | Y | Y | Y | Y | Y |
| 30(1)(a) | Data Protection Officer | Y | No DPV Concept | None | | Y | Y | Y | Y | Y | Y |
| 30(1)(a) | Representative | Y | No DPV Concept | None | | | Y | Y | Y | | Y |
| 30(1)(a) | Joint Controller | Y | dpv:DataController | Partial | | | Y | Y | Y | | Y |
| 30.1 | Business Process | N | dpv:PersonalDataHandling | Partial | | Y | Y | Y | | Y | |
| 30.1 | Owner of Process | N | dpv:DataController | Partial | | Y | | Y | | | Y |
| 30.1(b) | Purposes of processing | Y | dpv:Purpose | Exact | 65 / 33 | Y | Y | Y | Y | Y | Y |
| 6.1 | Legal Basis for Processing | N | dpv:LegalBasis | Exact | 6 / 6 | Y | Y | | | Y | Y |
| 30 (a) | Type of Processing | N | dpv:Processing | Exact | 9/ 33 | Y | | | | | |
| 30.1(c) | Categories of personal data | Y | dpv:PersonalDataCategory | Partial | 80/163 | Y | Y | Y | Y | Y | Y |
| 9.1 | Special Category Personal Data | N | dpv:SpecialCategoryPersonalData | Partial | 8/8 | Y | | | | | |
| 30.1(c) | Categories of data subjects | Y | dpv:DataSubject | Exact | 0 / 5 | Y | Y | Y | Y | Y | Y |
| 9.1 | Vulnerable Data Subject Category | N | dpv:DataSubject | Partial | | Y | | | | | |
| - | Classification Level | N | dpv:TechnicalOrganisationalMeasure | Partial | | Y | | | | | |
| 30.1(f) | Retention/Deletion Periods | Y | dpv:StorageDuration, dpv:StorageDeletion | Exact | | Y | Y | Y | Y | Y | Y |
| 6/14/30.1(b) | Data Combination | N | dpv:Combine | Exact | | Y | | | | | |
| 5.1 | Original Source of data | N | No DPV Concept | None | | Y | | | | | Y |
| 28 | Processor | N | dpv:DataProcessor | Exact | | Y | | | | | |
| 28.3 | Data Processing Agreement | N | dpv:Contract | Partial | | Y | Y | | Y | Y | Y |
| | Data Transfer | N | dpv:Transfer | Exact | | Y | | | | | |
| 28/30.1(c) | Data Categories subject to transfer | N | dpv:PersonalDataHandling, dpv:Transfer, dpv:PersonalDataCategory | Complex, Partial | | Y | | | | | |
| 30.1(d) | Categories of recipients of transfer data | Y | dpv:Recipient | Exact | 12/3 | Y | Y | Y | Y | Y | Y |
| 30.1(e) | Third countries that personal data are transferred to | Y | dpv:location | Complex, Partial | | Y | Y | Y | Y | Y | Y |
| 44-47 | Nature of Transfer to Third Country | N | dpv:LegalBasis | Partial | | Y | | | | | |
| 30.1(e) | Appropriate Safeguards for Third Country Transfers, | Y | dpv:TechnicalOrganisationalMeasure | Partial | | Y | | Y | Y | | Y |
| 32 | Technology Used | N | dpv:TechnicalOrganisationalMeasure | Partial | | Y | | | | | |
| 35 | Risk and Mitigation Measures | N | dpv:TechnicalOrganisationalMeasure, dpv:RiskManagementProcedure. | Complex, Partial | | Y | | | | | |
| 35 | Risk - Information about the risk | N | No DPV Concept | None | | Y | | | | | |
| 30.1(g) | Technical and organizational measures of security | Y | dpv:TechnicalOrganisationalMeasure | Exact | | Y | Y | Y | Y | Y | Y |
| 35 | Data Protection Impact Assessment | N | No DPV Concept | None | | Y | | | | Y | Y |
| 13/14/15 | Data Subject Rights | N | No DPV Concept | None | | Y | Y | | | | Y |
| 13 | Privacy Notice | N | No DPV Concept | None | | | | | | | Y |
| 6.1(f) | Legitimate interests for the processing | N | dpv:LegalBasis | Partial | | | | | | | Y |
| 6.1(f) | Legitimate Interest Assessment | N | dpv:LegalBasis | Partial | | | | | | | Y |
| 22.1 | Automated decision-making | N | dpv:Processing | Exact | | | | | | | Y |
| 6.1 | Link to record of consent | N | dpv:consent | Exact | | | | | | | Y |
| 5 | Location of personal data | N | dpv:StorageLocation | Exact | | | | | | | Y |
| 30.1 | Status of processing | N | dpv:PersonalDataHandling | Partial | | Y | | Y | | | |
| 33.5 | Personal Data Breach | N | No DPV Concept | None | | | | | | | Y |
| 30.1(f) | Retention and erasure policy. | N | TechnicalOrganisationalMeasure StorageRestriction | Exact | | | | | | | Y |
| 36.1 | Prior Consultation with DPA | N | No DPV Concept | None | | | Y | | | | |
| 30.1(b) | Main or Auxiliary Processing activity | N | Purpose | Partial | | | Y | | | | |

We identified that most ROPA's did not suggest any properties for user input. We identified that only 7 of the 43 GDPR concepts specified any properties on any ROPA's. These properties were matched against the DPV. The results are displayed in Table 1 in the column titled "Combined No. of Specified Field Values vs DPV". The DPV has the necessary expressiveness to meet these properties with the exception of purposes of processing, where the DPV will require additional properties.

## Conclusion

A ROPA document is a rich source of the personal data processing activities carried out by an organisation. We identified a set of six English language ROPA templates published by EU data protection regulators which contained a wide variation, scope and granularity for representing similar information. We identified the 43 unique GDPR concept fields across the six templates. This enabled the construction of the most comprehensive domain model of a ROPA to date that included both common and unique ROPA terms.

We mapped the 43 GDPR concepts identified using the DPV. We identified that 14 input fields can be fully mapped, 15 can be partially mapped, 3 can be mapped using a complex corresponding match, whilst 11 GDPR concepts cannot be matched as the DPV does not have sufficient expressive power.

This semantic analysis of the ROPA domain is the first step to developing a comprehensive ontology of ROPAs and information processing that will serve as the basis for intelligent GDPR compliance tools that support machine inference, data federation and integration. We have engaged with the DPVCG to incorporate our analysis towards representing ROPAs using the DPV. Our semantic model is a crucial component in this process by providing an indication of necessary fields to map ROPA templates published by EU data protection regulators. In addition, our analysis and mapping with DPV concepts provides a clear indication of further work in developing the DPV towards representing ROPA's and its utilization in the GDPR compliance process. A completed ROPA map would be a very useful resource that could be used by organisations to link with privacy tools. These tools can be built to query the ROPA and identify areas of non-compliance, and remedy or mitigate accordingly, as part of the accountability framework for an organisation.

## ACKNOWLEDGMENTS

This work is partially supported by Uniphar PLC., and the ADAPT Centre for Digital Content Technology which is funded under the SFI Research Centres Programme (Grant 13/RC/2106) and is co-funded under the European Regional Development Fund.